\documentclass[]{cas-sc}
\usepackage[backend=bibtex,sorting=none,maxcitenames=1000,maxbibnames=99,style=numeric-comp]{biblatex}
\usepackage{graphicx}
\usepackage{epstopdf}
\usepackage{fontawesome5}

\usepackage{amsmath,amsthm, amssymb, float, bm}
\usepackage{placeins}
\usepackage{longtable,tabularx,booktabs}
\usepackage{overpic}
\usepackage{cancel}
\usepackage{fancyhdr}
\usepackage{soul}
\usepackage{booktabs}
\usepackage{enumitem}
\usepackage[most]{tcolorbox}
\usepackage[inkscapelatex=false]{svg}
\usepackage{marginnote}

\definecolor{C0}{HTML}{1F77B4}
\definecolor{C1}{HTML}{FF7F0E}
\definecolor{C2}{HTML}{2ca02c}
\definecolor{C3}{HTML}{d62728}
\definecolor{C4}{HTML}{9467bd}
\definecolor{C5}{HTML}{8c564b}

\ifdefined\usetodonotes
\usetikzlibrary{calc}
\usepackage[textsize=scriptsize]{todonotes}
\usepackage{xparse}
\NewDocumentCommand{\brandon}{O{} m}{\todo[color=C0!10,linecolor=C0,bordercolor=C0,#1]{BR: #2}}
\newcommand{\np}[1]{\todo[color=C1!10,linecolor=C1,bordercolor=C1]{NP: #1}}
\usepackage{atbegshi} 
\AtBeginShipout{%
    \AtBeginShipoutUpperLeft{%
        \begin{tikzpicture}[overlay,remember picture]
            \fill [gray!25] (current page.north east) rectangle
                ($ (current page.south east) - (3cm,0cm) $);
        \end{tikzpicture}
    }
}
\else
\newcommand{\brandon}[1]{}
\newcommand{\np}[1]{}
\newcommand{\tododone}[1]{}
\fi

\ifdefined\usechanges
\usepackage{fontawesome}
\usepackage[commentmarkup=todo,authormarkup=none]{changes} 
\definechangesauthor[color=C0]{R1} 
\definechangesauthor[color=C1]{R5} 
\definechangesauthor[color=C2]{RA} 
\makeatletter
\@namedef{Changes@AuthorColor}{C2}
\colorlet{Changes@Color}{C2}
\renewcommand{\Changes@Markup@comment}[3]{%
  \IfStrEq{\Changes@optioncommentmarkup}{todo}%
		{\colorlet{Changes@todocolor}{authorcolor}\todo[color=Changes@todocolor!10, bordercolor=Changes@todocolor, linecolor=Changes@todocolor!70, nolist]{\textbf #1}}{}}
\makeatother
\fi

\makeatletter
\def\ps@pprintTitle{%
  \let\@oddhead\@empty
  \let\@evenhead\@empty
  \def\@oddfoot{\reset@font\hfil\thepage\hfil}
  \let\@evenfoot\@oddfoot
}
\makeatother

\usepackage[noabbrev,capitalise]{cleveref} 
\usepackage{subcaption} 

\makeatletter                                                                   
\newlength{\bibsep}{\@listi \global\bibsep\itemsep \global\advance\bibsep by\parsep} 
\makeatother

\everymath{\rm}

\begin{document}

\shorttitle{}
\shortauthors{Pushkar {\it et al}}
\title[mode=title]{Inverse method for determining general molecular weight distribution from polymer rheology.}

\author[isu]{Nihal Pushkar}
\author[uccs]{Xin C. Yee}
\author[uccs]{Jena McCollum}
\author[isu]{Brandon Runnels}[orcid=0000-0003-3043-5227]
\cormark[1]
\cortext[1]{Corresponding author}
\address[isu]{Department of Aerospace Engineering, Iowa State University, Ames, IA,  USA}
\address[uccs]{Department of Mechanical and Aerospace Engineering, University of Colorado, Colorado Springs, CO, USA}

\begin{abstract}
  Determination of polymer molecular weight distribution (MWD) from rheological measurements is desirable due to the ease and low cost of rheometry compared to other methods such as gel permeation chromatography.
  However, relating MWD to rheology requires the inversion of rheological models, for which there is no analytic solution.
  Prior approaches assume a functional form for the MWD (such as a lognormal or generalized exponential distribution), minimizing the error with respect to the functional form's degrees of freedom.
  While this is a powerful and robust technique for determining general polymer properties, such as average MWD or polydispersity, it requires former knowledge of the shape of the MWD.
  This work presents a generalized approach to solving the inverse problem directly, with no former knowledge of the MWD or assumptions regarding its functional form.
  To close the inverse problem and establish uniqueness, Lagrange multipliers constraints on the MWD are included.
  The method is applied with reptation-based models to a variety of polycarbonate, polyethylene and polystyrene polymers.
  For samples whose rheology are well-described by reptation, the predicted MWD is shown to match experimental measurements very well.
  For samples that are not well-described by reptation, the predicted MWD naturally differs from experiment.
  Nevertheless, the results still offer insight into how reptation-described polymers differ from their counterparts.
  This establishes the proposed inverse method as a viable practical tool for rheology-based characterization.
\end{abstract}


\maketitle

\section{Introduction}

Chain relaxation governs the viscoelastic response in most polymeric systems. 
Various structural features, such as molecular weight, molecular weight distribution (MWD), branching, and crosslinking, affect the spectrum of relaxation times that drive chain relaxation.
Linear polymers, in particular, are ubiquitous in many applications, making comprehensive characterization essential to ensure predictive and reliable performance. 
The determination of MWD is particularly important, as it controls rheological properties, density, and mechanical response. 
The primary method for determining MWD is gel permeation chromatography (GPC) \cite{cheremisinoff2001condensed} which, though accurate,  may be prohibitively expensive or even infeasible \cite{mandal2023review}. 
Rheology-based characterization of MWD is an attractive alternative to GPC \cite{ferry1970viscoelastic}, since the relaxation behavior of linear polymers is primarily determined by molecular weight distribution.
Rheology is cheaper, faster, and more robust (applicable to both soluble and insoluble polymers in common solvents), and is fairly sensitive to high molecular-weight constituents \cite{guzman2005regularization, shanbhag2010analytical}.
However, current methods for inferring MWD from rheology are limited.

Determination of rheology from MWD (the ``forward problem'') is well-established.
Numerous rheological models exist that connect the molecular weight distribution to rheological moduli, most of which fall into one of two main categories.
The first treats the polymeric system as a collection of entangled and cross-linked threads \cite{doi1988theory}, modeling chain dynamics through a tube-like control volume (reptation) over other tubes or tube-like structures.
These polymer chains are then represented as entangled Gaussian chain systems. 
The second approach models polymer chains as spring-and-dashpot systems, where the viscoelastic behavior arises from entanglements and cross-linking \cite{friedrich2009relaxation, pearson1978effect, das2006computational}.
In this case, the system is often modeled as a combination of Newtonian fluids.
Rheology-based characterization aims to leverage the accuracy of forward models by inverting them to predict MWD based on moduli data.

However, determining the MWD from rheological data (the "inverse problem") poses significant practical challenges.
While forward models reliably predict rheological behavior from a known MWD, inverting this relationship is inherently complicated or even impossible in closed form.
Rheological models that accurately describe polymer systems, such as those based on reptation theory or spring-dashpot analogs, are typically nonlinear and lack closed-form analytical solutions. 
This mathematical intractability is compounded by the ill-posed nature of the inverse problem: distinct MWDs can produce identical rheological responses within experimental uncertainty, leading to non-unique or multiple possible solutions \cite{touloupidis2023unified}.
For example, a bimodal MWD with well-separated peaks might yield the same storage modulus ($G'$) as a broad, unimodal distribution, making definitive conclusions impossible without additional constraints.
It is possible to regularize the solution by restricting the solution space (e.g. \cite{thimm1999determination}), or to use a Laplace transform, or to adopt a Laplace transform-based approach to extract the MWD analytically. 
Nevertheless, this approach is mathematically precise, requiring rheological data spanning the entire frequency spectrum-a condition rarely achievable experimentally due to instrumental limitations \cite{bersted1977relationship}. 
Such impractical requirements highlight the need for inversion techniques that are robust to finite or incomplete datasets.

In the absence of a direct mathematical inversion of forward models, one of the most successful approaches has been to recast variationally \cite{vanruymbeke2002determination, chaudhari2020rheological}, essentially using the method of least squares to obtain a reduced-order MWD in terms of a predetermined functional form (such as the generalized exponential, lognormal, Flory-Schultz, etc.) \cite{taletskiy2018entangled}. 
By non-dimensionalization, the number of free parameters are reduced further while ensuring that the crossover point (the frequency at which the loss and storage moduli are equal) is captured correctly \textcite{guzman2005regularization}. 
The empirical relation, $\tau \propto M^{\alpha}$ relates the relaxation time to the chain length of polymers in viscoelastic regimes \cite{lee2021calculation, serra2019viscoelastic, den2006viscosity}, closing the model.
A generalized relation between MWD and the relaxation time spectrum can then be developed, further improving the generality of MWD predictions \cite{nobile2008generalized}. 
Such methods can accurately estimate the MWD for some polymers, and has been validated against GPC measurements, as long as the the number of modes in the MWD is known and the polymer is well-described by the corresponding rheological theory.
However, many polymer systems exhibit complex or unknown MWDs, making them unsuitable for analysis using such reduced-order approximations, especially in polymer blends with an unknown number of constituents \cite{utracki2011rheology, paul1980polymer}.
Within a viscoelasticity-motivated framework, numerous models have been proposed \cite{lee2021calculation, serra2019viscoelastic}. 
Building on this foundation, reptation-based inverse methods have been developed \cite{den2006viscosity}.
Since viscoelastic polymer systems inherently exhibit multiple relaxation times, characterizing these systems necessitates a relaxation time spectrum \cite{monaco2022regularization, borg2009linear, monaco2009genetic}.
With rare exceptions (e.g. the Bayesian process presented by \textcite{shanbhag2010analytical}), inverse approaches require {\it a priori} knowledge of the MWD functional form.
This is obviously problematic if the form of the MWD differs considerably from the guessed functional form, and may produce incorrect of misleading statistics. 
Moreover, it is not necessarily possible to ascertain the degree of error induced by the functional form approximation, potentially leading to undetected discrepancies in the predicted MWD statistics.
Therefore, there remains a clear need for a robust inverse methodology for rheology-based characterization.

This work proposes a generalized inverse method for estimating the MWD from rheological behavior, with no restrictions on the functional form of the MWD or the type of forward rheological model. 
An optimization approach is proposed similar to prior work; however, instead of restricting to a fixed-form parameterization, the optimization space is expanded to include all probability distributions.
Essential constraints on the MWD - non-negativity and unitarity - are replaced by boundary value constraints and penalty terms, respectively, imposed on the cumulative distribution.
Additionally, an optional smoothness condition is included that improves the regularity and uniqueness of the solution.

The remainder of the paper is structured as follows. 
The salient forward models are reviewed, and the constrained inverse optimization problem is formulated (\Cref{sec:theory}).
Next, the proposed inverse method is applied to a variety of polymer systems to demonstrate its fidelity and robustness (\Cref{sec:results}).
Finally, the capabilities and limitations of the proposed method are reviewed in the context of future directions (\Cref{sec:conclusion}).

\section{Theory}\label{sec:theory}

\subsection{Determining rheological behavior from molecular weight distribution}

Here we briefly review classical models for polymer rheology in order to standardize notation and set the context for the proposed inverse method.
Most rheological models aim to relate the molecular weight distribution, (MWD, $W(M)$) to the rheological properties (storage modulus $G'(\omega)$ and loss modulus $G''(\omega)$) in an averaged, or statistical sense.
To avoid confusion, it is clarified that this work always defines distribution such as $W(M)$ to be normalized with respect to the log measure over $(0,\infty)$; that is,
\begin{align}
  |W| = \int_{-\infty}^{\infty} W(M)\,d(\ln(M)) = \int_0^\infty W(M)\,\frac{dM}{M}  = 1, \label{eq:W_unitary}
\end{align}
which renders $|W|$ invariant with respect to scaling factors of $M$, and coincides with the area norm as represented in log space.
For simple, constrained systems, it is possible to derive exact analytical expressions for these relationships \cite{shanbhag2012analytical} in terms of a kernel that relates the MWD to the constitutive behavior of the polymer.
Forward models generally fall into two main categories.
The first category treats the polymeric system as a system of entangled chains (e.g. \cite{doi1988theory,cloizeaux1988double}), relating the complex modulus to MWD via
\begin{align}
    G_t(t) = G_N \bigg( \int_{M_C}^{\infty} W(M) F(\tau(M), t)^{\frac{1}{\beta}} \frac{dM}{M}\bigg)^{\beta},
    \label{eq:Gt_gaussianchains}
\end{align}
where $G_t$ is the time-domain modulus, the kernel $F$ is the real space transfer function for the relaxation of a viscoelastic system with relaxation time $\tau$, and $G_N$ is a material parameter.
The parameter $\beta$ describes the degree of reptation, so that $\beta=1,2$ corresponds to the special cases of single and double reptation, respectively.
(It is noted that, though this work primarily considers reptation models, it is generalizable to arbitrary rheological models.)

The second category of models treat polymer rheology phenomonenologically as a system of elementary mechanical elements (springs and dashpots) \cite{cho2016viscoelasticity}, which condenses polymer behaviors into a simple relaxation time \cite{stadler2011understanding,vanruymbeke2007quantitative}.
This allows the relaxation times scales of the material to be described via a ``relaxation spectrum'' ($h$), characterizing how perturbations decay/grow with time.
This is then related to the rheological response by the relation \cite{ferry1970viscoelastic},
\begin{align}
    G(t) = \int_{\tau = 0}^{\infty} h(\tau)\, e^{-t/\tau} \frac{d \tau}{\tau}.
    \label{eq:ferry-G}
\end{align}
The modulus equation (\cref{eq:ferry-G}) can then be transformed into reciprocal space in order to obtain the storage and loss moduli \cite{ferry1970viscoelastic},
\begin{align}
  G'(\omega) &= \int_0^\infty h(\tau) \frac{(\omega\tau)^2}{1 + (\omega\tau)^2}\frac{d\tau}{\tau}
  &
    G''(\omega) &= \int_0^\infty h(\tau) \frac{(\omega\tau)}{1 + (\omega\tau)^2}\frac{d\tau}{\tau},
                  \label{eq:Gp_Gpp}
\end{align}
The above integrals are problematic to compute because $h(\tau)$ is not necessarily known or characterized.
To estimate $h$, it may be assumed that the relaxation time is related to the effective polymer chain length; for example, \textcite{tsenoglou1991molecular} proposed 
\begin{align}
  \tau \approx k M^\alpha,
\end{align}
where $k,\alpha$ are material parameters (temperature dependent) corresponding to a reference relaxation time and the degree of 0-shear dependence on molecular weight, respectively.
Such an approximation enables the time-domain integral to be written in $N$-integral form (after non-dimensionalizing M).
This leads to the simplifying approximation $\tau \propto N^{\alpha}$, that results in the following simplified expression
\begin{align}
  G'(\Omega) &= \int_0^\infty H(N) \frac{(\Omega N^\alpha)^2}{1 + (\Omega N^\alpha)^2} \frac{dN}{N}
  &
  G'(\Omega) &= \int_0^\infty H(N) \frac{(\Omega N^\alpha)}{1 + (\Omega N^\alpha)^2} \frac{dN}{N},
\end{align}
where $\alpha$ is the measure of the degree of zero-shear dependence of the polymer on its molecular weight.
The dimensionless molecular weight $N$ and dimensionless frequency $\Omega$ are defined as
\begin{align}
  &
  m_p = \int_{M=0}^{\infty} M W(M) d(\ln{M})
  &
  N = \frac{M}{m_p} &
  &
    \Omega = \omega \tau_0,\label{eq:relaxation_spectrum}
\end{align}
in which $m_p$ is the characteristic molecular weight.
Then a modified relaxation spectrum proposed in \cite{thimm1970analytical}, which defines the general repation relaxation time spectrum, in which $\beta=2$ (double reptation) is used to relate nondimensional rheological relaxation time $H$ to dimensionless MWD, denoted $\overline{W}(N) = W(m_pN)$
\begin{align}
  &
  H(N) = \beta \overline{W}(N) \left( \int_{N}^{\infty} \frac{\overline{W}(N')}{N'} dN' \right)^{\beta - 1}
  \label{eq:define_HN}
  & H(N) = 2 \overline{W}(N) \int_N^\infty \overline{W}(N')\,\frac{dN'}{N'}.
\end{align}
It can then be shown using the definitions of $H(N)$ (\cref{eq:define_HN}) that $|H|= \int || H(N)/N || dN = 1$.
Alternately, $H(N)$ can be written as
\begin{align}
  H(N) &= -N \frac{d}{dN} P(N)^2
  \label{eq:special_PN-HN}
  &
    P(N) = \int_{N}^\infty \overline{W}(N')\,\frac{dN'}{N'},
\end{align}
where $P$ is the cumulative MWD; this form is convenient for the proposed inverse method discussed in this paper.
The expressions for complex moduli (\cref{eq:Gp_Gpp}) along with the model for relaxation spectrum (\cref{eq:relaxation_spectrum}) constitute a closed rheological forward model.

\subsection{Solving the inverse problem for arbitrary molecular weight distribution}

As discussed above, direct inversion of forward models for polymer rheology is not possible in general, because the mapping for an arbitrary model is neither injective (multiple MWDs can produce the same rheological behavior) nor surjective (some rheological behaviors exist for which there is no corresponding MWD under a given model).
An implicit, approximate approach is required in which certain restrictions are placed on $W(M)$ in order to render the solution unique and the problem well-defined.
Given a set of experimental rheological data $(G_e',G_e'')$ a least-squares-type solution may be approximated by the solution to
\begin{align}
  W(M) = \underset{W(M)}{\operatorname{arg\ inf}} \Big[E_1(G_e',G_m'[W(M)]) + E_2(G_e'',G_m''[W(M)])\Big]
  \ \ \ \
  \text{subject to constraints on } W(M),\label{eq:general_optimization}
\end{align}
where $E_1,E_2$ are error metrics, and the bracketed argument $[W]$ indicates functional dependence on $W$.
The solution is then dependent on a robust choice of $E_1,E_2$, and suitably minimal restrictions on $W$.
A natural choice of error is derived from the modification of the $L_2$ norm under a log measure $\mu$,
\begin{align}
  E(G_e,G_m) = \int_{-\infty}^\infty (\ln G_{\text{m}}(\omega) - \ln G_{\text{e}}(\omega))^2\,\,d\ln\omega,
\end{align}
so that the degree of error corresponds to the intuitively meaningful geometric area of the difference in log-log space.
\textcite{guzman2005regularization} proposed the form for error based on a semi-norm, which we write here an a more suggestive (yet equivalent) form:
\begin{align}
  E(G_e,G_m) &= \int_{0}^{\infty}\bigg[\ln\bigg(\frac{G_m(\omega/\omega_{mC})}{G_{mC}}\bigg) - \ln\bigg(\frac{G_e(\omega/\omega_{eC})}{G_{eC}}\bigg)\bigg]^2\,d\mu(\omega)
  &
    \mu = \sum_i\delta(\omega-\omega_i),
\end{align}
where $\delta$ is the Dirac delta distribution, and the crossover point coordinates $(\Omega_{mC},G_{mC})$, $(\Omega_{eC},G_{mC})$ are predicted by the model and measured by experiment, respectively.
The model-predicted crossover point $(\Omega_{mC},G_{mC})$ is itself a functional on $W$.
The measure $\mu$ reduces the integral to a sum over discrete experimental points $(\Omega_i,G'_{ei},G''_{ei})$ enumerated by $i$, so that the weight of the error is determined by the density of experimental data points rather than an objective measure.

Two factors make \cref{eq:general_optimization} difficult to solve.
First is the lack of convexity of $E_1+E_2$ with $W(M)$; the functional dependence is general, nonlinear, and arbitrarily complex.
This can induce numerical challenges in practical computation, making the determination of a global solution problematic.
The second difficulty are the constraints placed on $W$, which are:
\begin{align}
  &\text{non-negativity: }W(M) \ge 0
  &
  &\text{unitarity: } |W(M)| = \int_0^\infty W(M)\,\frac{dM}{M} = 1,
\end{align}
Prior approaches adopt a functional form for $W=f(\mathbf{q})$ (where $\mathbf{q}=(q_1,q_2,\ldots)$ are a generalized set of degrees of freedom for $f$) that is guaranteed to satisfy unitarity and non-negativity for all possible values of $\mathbf{q}\in\mathbb{R}^N$.
This allows the constrained, infinite-dimensional optimization problem to be replaced with the unconstrained problem,
\begin{align}
  \mathbf{q}
  = \underset{\bm{\mathbf{q}}\in\mathbb{R}^N}{\operatorname{arg\ inf}} \Big[E_1(\mathbf{q}) + E_2(\mathbf{q})\Big],
\end{align}
which is significantly more tractable.
Even with fairly complex functional dependencies of $E_1+E_2$ on $\mathbf{q}$, the unconstrained finite-dimensional optimization problem is readily solvable on modern computers.
However, without prior knowledge of the form of $W$, it is not possible to know whether the predicted MWD is accurate; a bimodal distribution, for example, cannot be suitably modeled with a GEX or LN functional form.
Moreover, there is no clear method of determining the sensitivity of the solution to the choice of functional form, significantly limiting the method.

\subsubsection{Imposing unitarity though boundary conditions}

The unitarity constraint on $W$ (\cref{eq:W_unitary}) is unwieldy to implement because it is a nonlocal (i.e. requires an integral to enforce). 
While unitarity can be enforced (for example) using Lagrange multipliers, doing so adds additional complexity to the model and reduces interpretability.
However, it is straightforward to show that the unitary constraint on $\overline{W}(N)$ is equivalent to
\begin{align}
  P(0) &= 1
  &
    \lim_{N\to\infty} P(N) = 0,
\end{align}
Or, if the MWD can be bound to a known interval $[N_a,N_b]$, then, simply
\begin{align}
  P(N_a) &= 1  &   P(N_b) &= 0.
\end{align}
Since double reptation allows for the forward model to be written in terms of $P(N)$, then the unitary constraint reduces to a boundary value constraint on $P(N)$, which is far easier to implement: all that is necessary is to restrict the optimization space to the interior of the interval $(N_a,N_b)$.
Unitarity is then guaranteed by construction, and no explicit constraints are necessary.

\subsubsection{Imposition of non-negativity}

Equally important is the constraint of non-negativity of $\overline{W}(N)$, which can be written in either of the following equivalent forms:
\begin{align}
  \overline{W}(N) \ge 0  \ \ \ \ \ \Leftrightarrow \ \ \ \ \  \frac{dP}{dN} \le 0.\label{eq:decreasing_constraint}
\end{align}
The constraint, written in terms of negativity on the derivative of $P$ (insuring predicted $W$ is non-negative), is difficult to impose explicitly, even though the constraint itself is both local and linear.
To remedy this, a penalty method is proposed, based on the equivalent requirement that
\begin{align}
  \lim_{\mu \to \infty} \exp\Big(\mu \frac{dP}{dN} \Big) = 0  \ \ \  \mu \in \mathbb{R}^+,
\end{align}
where $\mu$ is a strictly positive penalty parameter that is similar in function to a Lagrange multiplier.
This can be thus expressed variationally as:
\begin{align}
    P \in \underset{P}{\operatorname{arg inf}}\ \Big[ \ \sup_{\mu>0} \ \ \int_{N_a}^{N_b} \exp\Big(\mu \frac{dP}{dN}\Big)\,dN\Big],
\end{align}
whence it is clear that the objective function returns $\infty$ unless $dP/dN$ is non-positive everywhere.
This form is useful because it can be readily combined with the original error function, and also because it provides a convenient means of regularizing the non-negativity constraint through a chosen parameter $\mu$.
Theoretically $\mu$ should be allowed to tend towards infinity; in practice, it suffices simply to choose a $\mu$ that is large enough to ensure compliance with the constraint up to tolerance.

\subsubsection{Smoothness penalty}

Even with the above constraints, the inverse problem still remains ill-posed due to the lack of uniqueness: for any given rheological behavior, there is not a unique MWD that minimizes the error - even accounting for the essential constraints.
One may see this intuitively by supposing that that an optimal MWD, $W^*$ has been found.
If a small, alternating perturbation with frequency $f$ is added to $W*$, one can observe that the effect on $H$ (\cref{eq:define_HN}) will be minimal as $f\to\infty$, even if the amplitude of the perturbation is large, because of the integral formulation of $H$.
Any residual effect will be further dampened in the calculation of $G',G''$, as these are also integral equations on $H$.
Therefore, when performing the inverse calculation, it is likely that that a highly noisy solution is obtained.
While this in itself is not necessarily problematic (MWD may certainly exhibit distinct peaks), it is nonetheless necessary to enforce controllable behavior of the inverse method.
Thus, the following supplementary constraint,
\begin{align}
  \bigg|\frac{d^2P}{dN^2}\bigg|  \le \xi,\label{eq:less_than_xi}
\end{align}
is added to ensure that the solution is reasonably smooth, where $\xi$ is a parameter of choice signifying the acceptable level of noise in the solution.
The inclusion of the smoothness constraint is also essential to ensure that the boundary conditions are properly enforced. 
By penalizing the second derivative, discontinuities at the edge of the domain (just above $N_a$, just below $N_b$) are intolerable.
This facilitates the propagation of the boundary conditions into the main solution.

Integrating the smoothness constraint of \cref{eq:less_than_xi} into the objective function, the modified optimization problem becomes
\begin{align}
  P = \underset{P}{\operatorname{arg} \inf} \lim_{\substack{\mu\to\infty , \lambda \to 0}}
  \Bigg[
  \underbrace{E_1( G_e',G_m'[P]) + E_2(G_e'',G_m''[P]) \Bigg.}_{\text{error norm} \chi^2}
  + 
  \underbrace{\lambda\int_{N_a}^{N_b}\bigg|\frac{d^2P}{dN^2}\bigg|\,dN }_{\text{smoothness}}
  + 
  \underbrace{\int_{N_a}^{N_b}\exp\Big(\mu \frac{dP}{dN}\Big)\,dN}_{\text{non-negativity}}
  \Bigg],\label{eq:penalty_method}
\end{align}
where the parameters $\lambda,\mu$ are penalty parameters corresponding to the non-negativity and smoothness constraints, respectively.
The limit of $\mu$ to $0$ indicates that a too-large value of $\mu$ compromises the actual error norm.

In practice, it is most practical to $\mu$ and $\lambda$ as finite parameters that can be chosen or adjusted depending on the problem.
For the non-negativity parameter $\mu$, one may begin with a fairly small value, until the solution converges (verifying that the non-negativity constraint is met to required precision, of course).
The choice of parameter $\lambda$ determines the degree of smoothness desired in the final solution.
It should be small enough to ensure that the objective function has been suitably minimized.
Given that the solution is non-unique, there may be different equivalent solutions (with different degrees of smoothness) that are determined by different choices of $\lambda$.


\section{Results and Discussion}\label{sec:results}

The generalized inverse method is tested on previously published data with behavior that is known to be well-described by reptation-based forward models.
Where applicable, forward model parameters are set to those in the literature where they are used.
Specifically, rheological data along with GPC measurements of polycarbonate (PC) and polystyrene (PS) are taken from \textcite{guzman2005regularization, vanruymbeke2002determination}.

The inverse method is implemented in python, where the objective function, initialization, solver and tolerance are defined.
To perform the unconstrained minimization, the BFGS (L-BFGS-B) solver implemented in SciPy \cite{virtanen2020scipy} is used.
Since the objective function is highly nonconvex, there can be a strong dependence of the optimized solution on the initial conditions, so they must be chosen with care to avoid erroneous solutions or poor solver performance.
A laterally inverted sigmoid function is used for $P(N)$, to avoid excessive initial penalization of discontinuities with tolerance of $10^{-1}$ for the Lagrange multipliers method.

An advantage of using the generalized method is the ability to consider convergence of the solution with respect to the number of $n$ points.
In each case, the number of points is increased from $\sim\mathcal{O}(10)$ to $\sim\mathcal{O}(10^3)$ (the exact values depend on the system) and the MWD for each resolution is reported and compared with GPC (experimental) measurements.
Due to the non-convexity of the objective function convergence to a single solution is not ensured.
It was observed that the random initialization for different resolution were converging on totally different solutions (which do show why this problem is ill-posed and can have multiple solutions).
To address this, successive refinement and interpolation is used to initialize starting points for each increased resolution.
As $N$ increases, the solution for the former resolution is interpolated to the new resolution and used as initial guess for increased resolution.
This improves the performance of the optimizer and ensures that the solution for each resolution is at or near the same minimum.

It is emphasized here that the results demonstrate the capacity of the inverse method, not necessarily the accuracy of the forward model.
For consistency, the same forward model is used for all polymers, but it is crucial to note that not all polymers are accurately modeled using double reptation.
When applying double reptation as a forward model, this presents as a mismatch between predicted and observed rheological data.
On the other hand, when applying double reptation as an inverse method -- as is the case here -- the discrepancy presents as a mismatch between observed MWD and predicted MWD.
That is, the inverse method demonstrates the requisite MWD to yield specific rheological data, assuming that the polymer is perfectly modeled by reptation.
However, as will be shown here, the predicted MWD for non-reptation-based polymers can yield insight into the mechanisms underpinning the rheological behavior.

\subsection{Polycarbonate}

Polycarbonate polymers have been well-described with reptation-based models, as observed in the literature, and are therefore ideal candidates for demonstrating the use of the inverse method proposed here. 
In fact, it is generally possible to capture salient characteristics of the MWD with a relatively small frequency range around the crossover point.
The goal of this section is to determine whether a general model, with an arbitrarily high number of degrees of freedom, is capable of capturing the MWD as well.

All PC characterization data was originally published in \textcite{vanruymbeke2002determination}.
Four PC datasets are considered here based on availability of MWD characterization.
The first (denoted PC-1) is a commercially available linear bisphenol made by interfacial phosgenation, which was supplied to the original authors by GE Plastics.
PC-1 had a polydispersity of 2.35.
The other three samples, PC-2, PC-3 and PC-4, are fractions of polymer PC-1 with polydispersities of 1.98, 1.52 and 1.39, respectively.
Although they are from a common source, the variance in polydispersity implies different mean masses, creating diversity in this dataset, making it possible to test the model's performance for various instances from a single source.
The original source suggests the parameters $\alpha = 3.60, \beta = 2.00$ describe the moduli optimally in the domain up to $20 \times \omega_C$ for PC-1.
For PC-2, PC-3 and PC-4, the literature values of $\alpha = 3.10, \beta = 2.25$ are used.

\begin{figure}
    \centering
    \includegraphics[width=\linewidth]{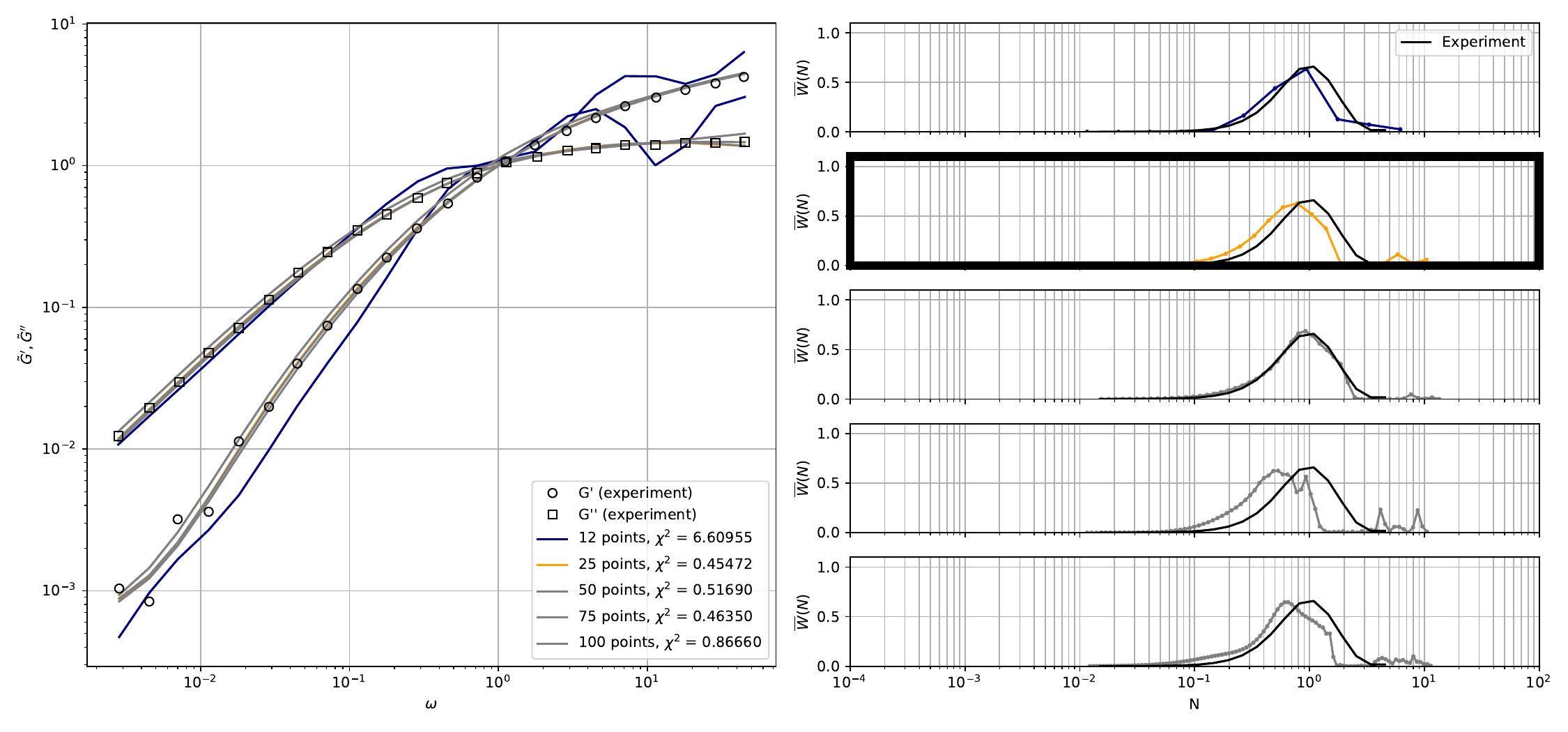}
    \caption{
    Inverse method results for PC-1. 
    Experimental moduli data from Fig 8a and MWD data from Fig 1 in \cite{vanruymbeke2002evaluation}.
    The bold box indicates the optimal fit to $\chi^2$ (local minima) which is obtained at $n = 25$ with parameters $\lambda = 3.0$, $\mu = 5.0$. 
    (Notably, $n=50$ best captures MWD, despite having a higher $\chi^2$ value.)
    When $n \geq 75$, there is no objective improvement in $\chi^2$ and is therefore discarded (gray). 
    Visible overfitting occurs at N$\ge$75.
    }
    \label{fig:PC1-final}
\end{figure}

For PC-1 (\cref{fig:PC1-final}), the moduli are captured with high precision for the LM method. 
After the first 1-2 refinements, the match to the moduli is nearly exact. 
The LM method (\cref{fig:PC1-final}) yields a reasonable estimate for $n=12$, even though the corresponding rheological predictions are very poor (except for the crossover point).
Increasing $n$ to $n=25$ significantly improves the match to both rheological data and the MWD, with nearly an exact match observed for both. 
To determine sensitivity, $n$ is increased by up to 100 points. 
As observed in similar cases, there is no significant improvement in the rheological match; instead, the optimizer attempts to capture higher-order fluctuations in the MWD (i.e., overfitting the modified $\chi^2$), resulting in the emergence of spurious peaks for higher $N$ values. 
In conclusion, the LM model performs well at predicting MWD, as long as $n$ is not increased beyond the resolution necessary to capture the rheological data.

\begin{figure}
    \centering
    \includegraphics[width=\linewidth]{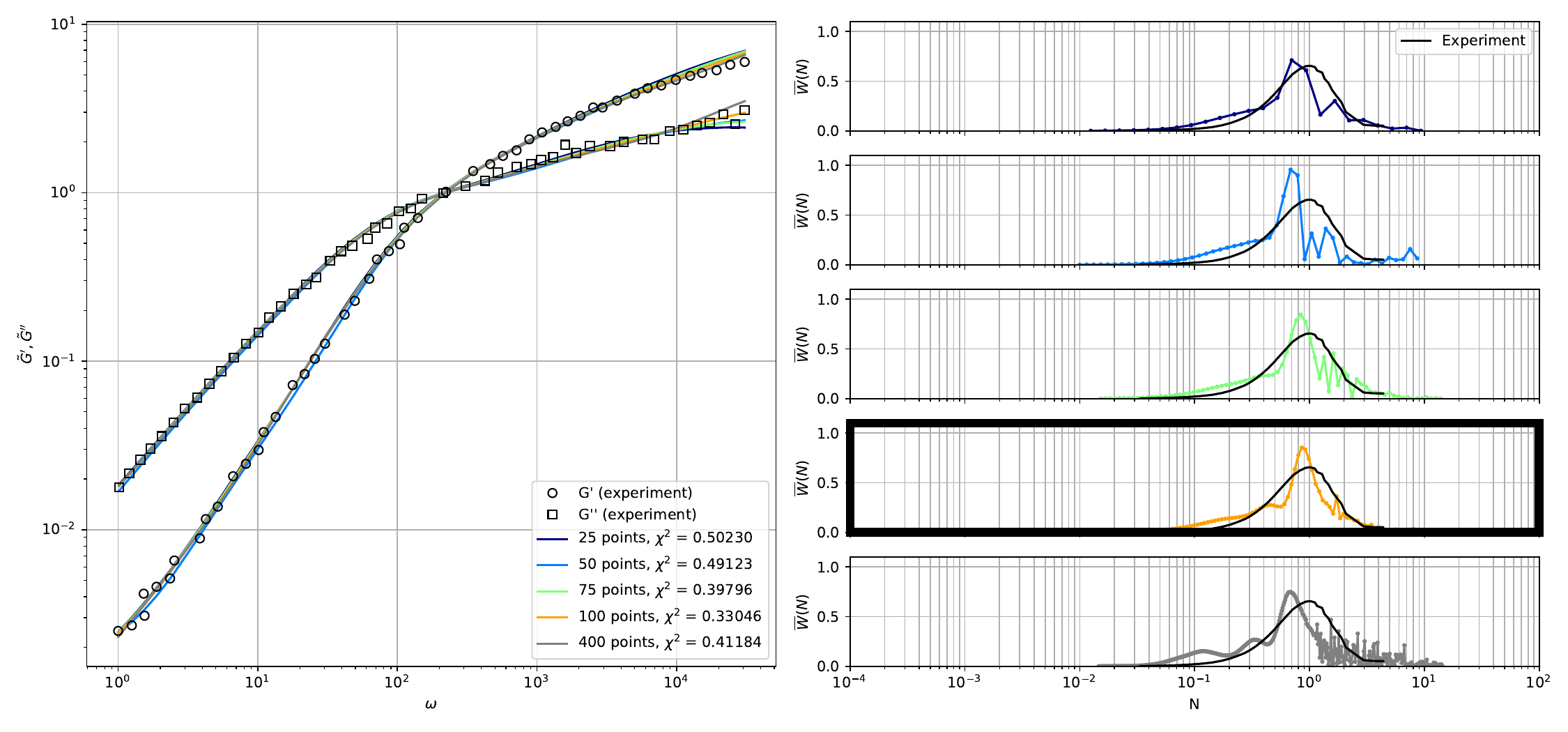}
    \caption{
    Inverse method results for PC-2. 
    Experimental moduli data is from Fig 7b  and MWD data from Fig 1 (MWD) of \cite{vanruymbeke2002evaluation}.
    The bold box indicates the selected optimal fit based on the $\chi^2$ criterion, above which overfitting occurs (indicated in gray).
    Regularization parameters were $\lambda = 7.0$, $\mu = 7.0$. 
    It is noted that an very good fit for the rheological data is attained, despite the discrepancy between predicted MWD, indicating deviation of the polymer from reptation-like behavior.
    }
    \label{fig:PC2-final}
\end{figure}

The PC-2 sample (\cref{fig:PC2-final}) is known to be less well-described than PC-1 by double reptation models due to the presence of Rouse modes at higher frequencies. 
The inverse method captures the rheological data with accuracy, with the only distinction observed for low values of $n$. 
A stable predicted MWD emerges at approximately $n=70$ and persists for increasing $n$ (until overfitting begins at $n=100$). 
This highlights two observations: (1) the need for constraints (e.g., smoothness) to find a unique solution, and (2) the inability of the forward model to accurately represent this polymer system, as the predicted MWD diverges from the actual data.

\begin{figure}
    \centering
    \includegraphics[width=\linewidth]{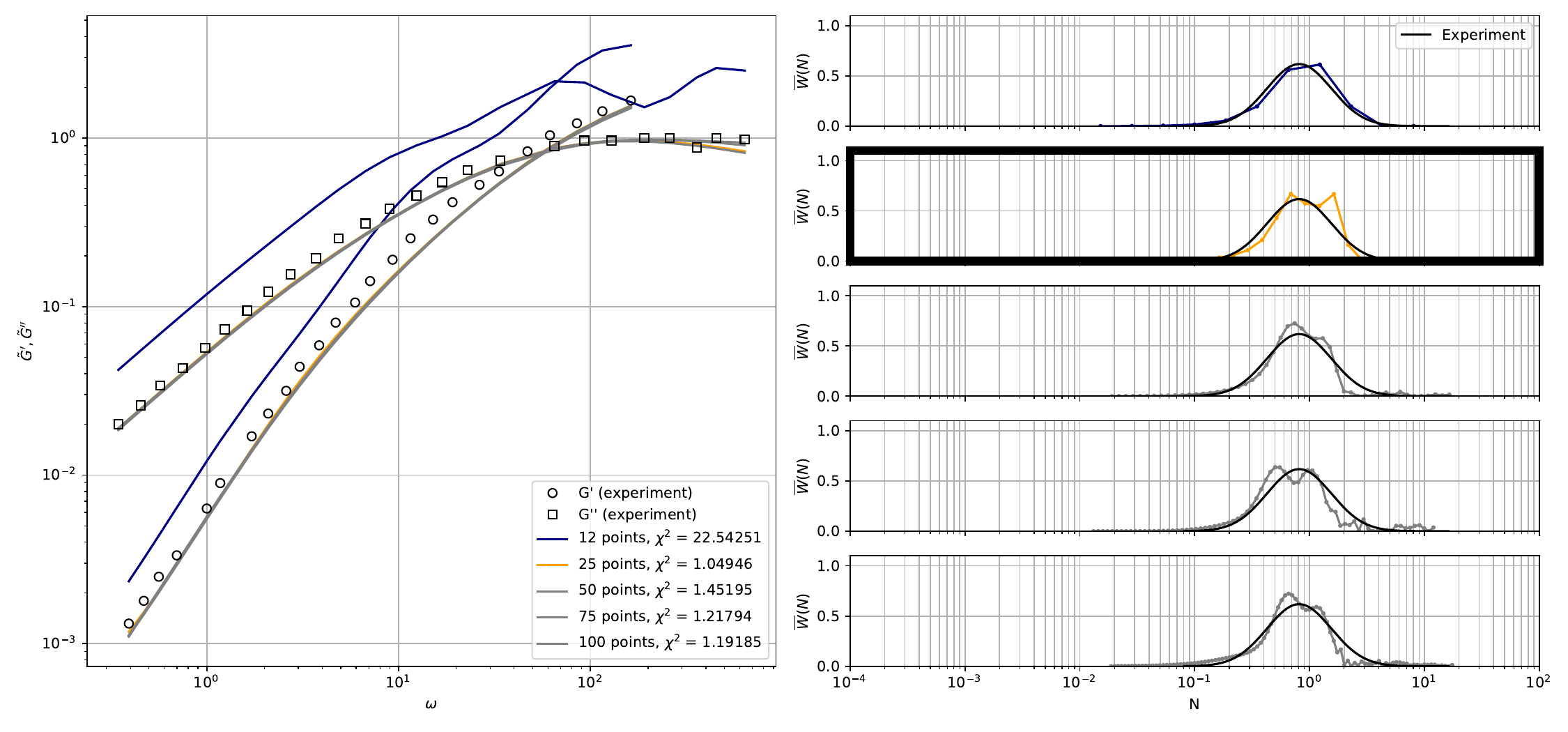}
    \caption{
    Inverse method results for PC-3.
    Experimental moduli data is from Fig 8b and MWD from and Fig 1 (MWD) of \cite{vanruymbeke2002evaluation}.
    The bold box indicates the optimal fit to $\chi^2$ (local minima) which is obtained at $n = 25$ with parameters $\lambda = 7.0$, $\mu = 7.0$. 
    Of note is the slight double peak in the predicted MWD, indicating a bimodal distribution of relaxation times.
    Otherwise, both the underresolved (n=12) and overfitted results (n>25) yield a good match both to rheological data and to MWD.
    }
    \label{fig:PC3-final}
\end{figure}

For PC-3 (\cref{fig:PC3-final}), the method converges almost immediately, with no appreciable changes in the objective function observed for $n>25$ (\cref{fig:PC3-final}).
Increasing $n$ from 25 to 100 results in little change in the MWD, giving confidence to the prediction. 
Overfitting is observed for $n>100$. The MWD consistent with the rheological data under the double reptation model is bimodal, with similar polydispersity index (PDI) and average molecular weight. 
While the MWD does not match the experiment, a reasonable estimate is still obtained, highlighting the need for a more robust forward model.

\begin{figure}
    \centering
    \includegraphics[width=\linewidth]{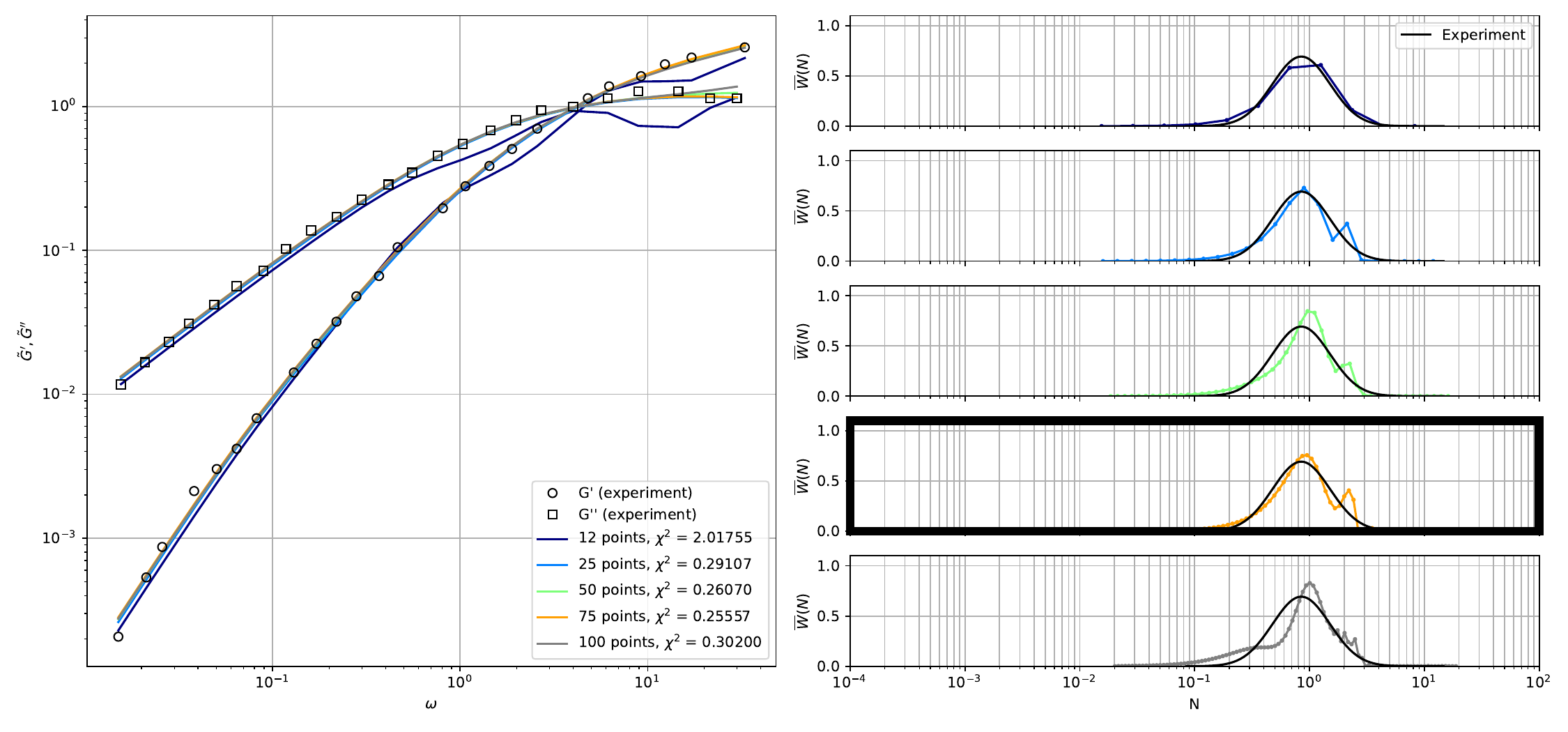}
    \caption{
    Inverse method results for PC-4.
    Experimental moduli data from Fig 8c and experimental MWD from Fig 1 of \cite{vanruymbeke2002evaluation}).
    The bold box indicates the optimal fit to $\chi^2$ (local minimum) which is obtained at $n = 75$ with parameters $\lambda = 7.0$, $\mu = 7.0$. 
    Because the constraints ensure smoothness, the MWD converges relatively well. 
    }
    \label{fig:PC4-final}
\end{figure}

The PC-4 sample (\cref{fig:PC4-final}) exhibits a close match for both MWD and rheology, indicating that it is well-described by the reptation model. 
The LM method yields a close match for all $n$, indicating a robust fit (\cref{fig:PC4-final}). 
A small secondary peak at $N=2.0$ emerges at $n=25$ and persists with increasing $n$. 
This reflects polymer behavior not fully captured by reptation, though the average MWD remains a close fit.
Notably, this sample shows less susceptibility to overfitting, possibly due to well-behaved rheological data or sampling frequencies.

\subsection{Polyethylene}

Reptation models and mixing rules have been used to predict relaxation behavior of other polymers including high-density polyethylene (PE).
The inverse method is tested on broad PE samples from literature: PE1, synthesized via Ziegler-Natta catalysis, and then fractionated into narrower molar mass distributions (PE2, PE3, and PE4) using successive solution fractionation \cite{vanruymbeke2002evaluation}. 
(It is noted that, since the crossover point is essential for the inverse method, the PE2 sample is excluded from this study.)
These PE datasets are well described by the highly sophisticated time-dependent diffusion with double reptation (TDD-DR) model, making them ideal for testing with spring-dashpot-type models. 
This approach not only challenges the limits of the macroscopic model but also examines how inverse methodologies respond to theories for which they were not originally developed.

A challenge in applying the inverse method to the rheological data for the PE datasets is the lack of high-frequency data.
While this is not a problem when considering rheological prediction from known MWD, it is problematic in determining the entire MWD range using the inverse method.
The results in this section demonstrate the increased variability in predicted MWD that results from the lack of high frequency data.
However, as will be shown, it is possible to achieve a very close match to rheological data in all cases.
To improve the MWD estimate, additional data (as well as a more advanced forward model) are required.
But even without these, the MWD estimate is found to be reasonable.

\begin{figure}
    \centering
    \includegraphics[width=\linewidth]{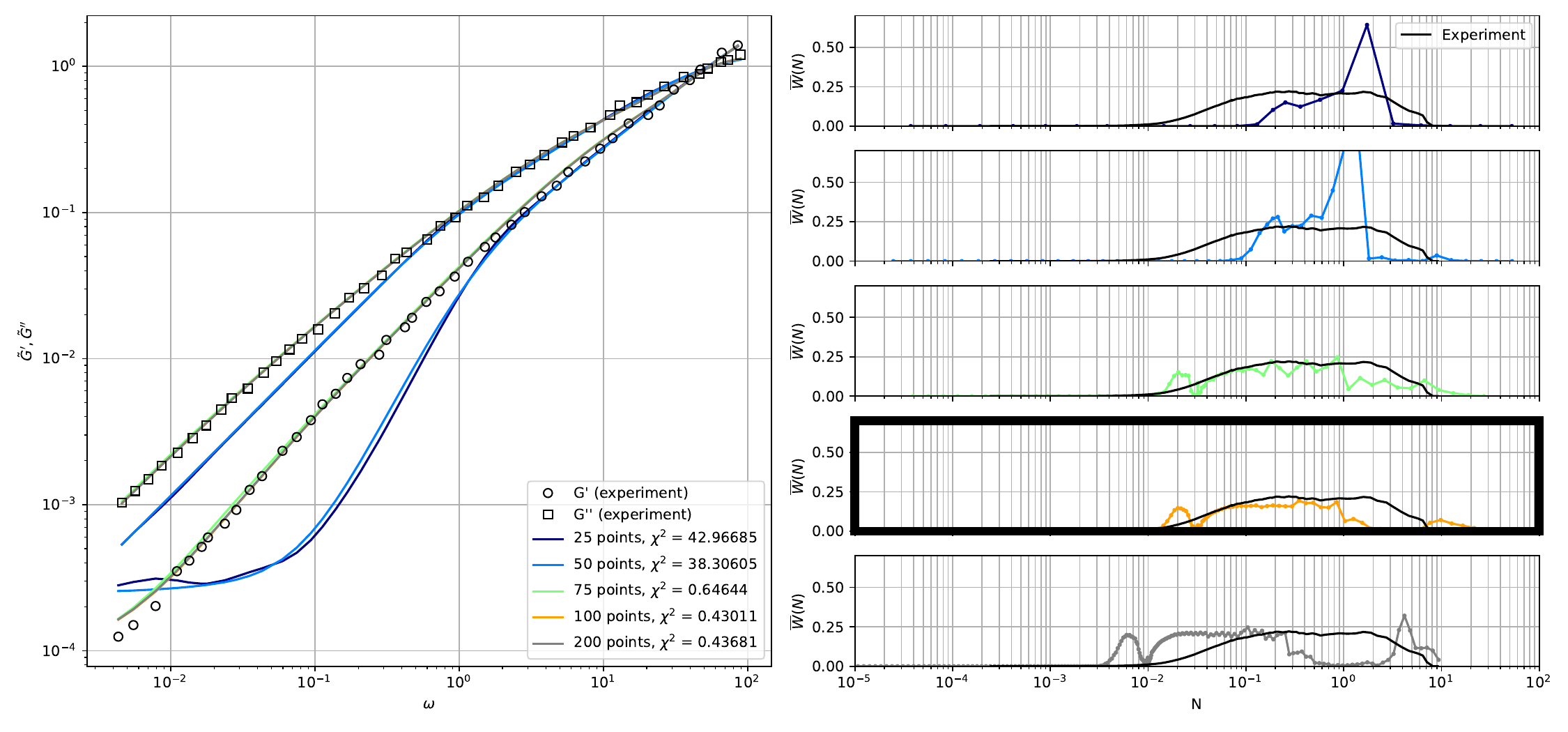}
    \caption{
        Lagrange multiplier method for PE1.
        Inverse method results for PE1.
        Experimental moduli data is from Fig 9a and experimental MWD from Fig 1b of \cite{vanruymbeke2002evaluation}
        The bold box indicates the optimal fit to $\chi^2$ (local minima) which is obtained at $n = 100$ with parameters $\lambda = 5.0$, $\mu = 5.0$.
        Increasing $n$ overfits the moduli curves, without significant improvement in MWD predictions. 
    }
    \label{fig:PE1-final}
\end{figure}

The method is applied first to the PE1 rheological data (\cref{fig:PE1-final}).
In both cases, it is clear that the rheological match is nearly exact (or at least up to standard deviation from the trend), in particular near the crossover point.
For low resolution MWD, the algorithm behaves relatively poorly, indicating complexity in the optimal MWD.
The criterion for selection is satisfied at n=100, after which overfitting begins to occur.
The predicted MWD is similar to the experimental measurement, and has mostly similar statistical properties, but exhibits a number of additional features such as small local peaks at the lower and higher end of its support.

\begin{figure}
    \centering
    \includegraphics[width=\linewidth]{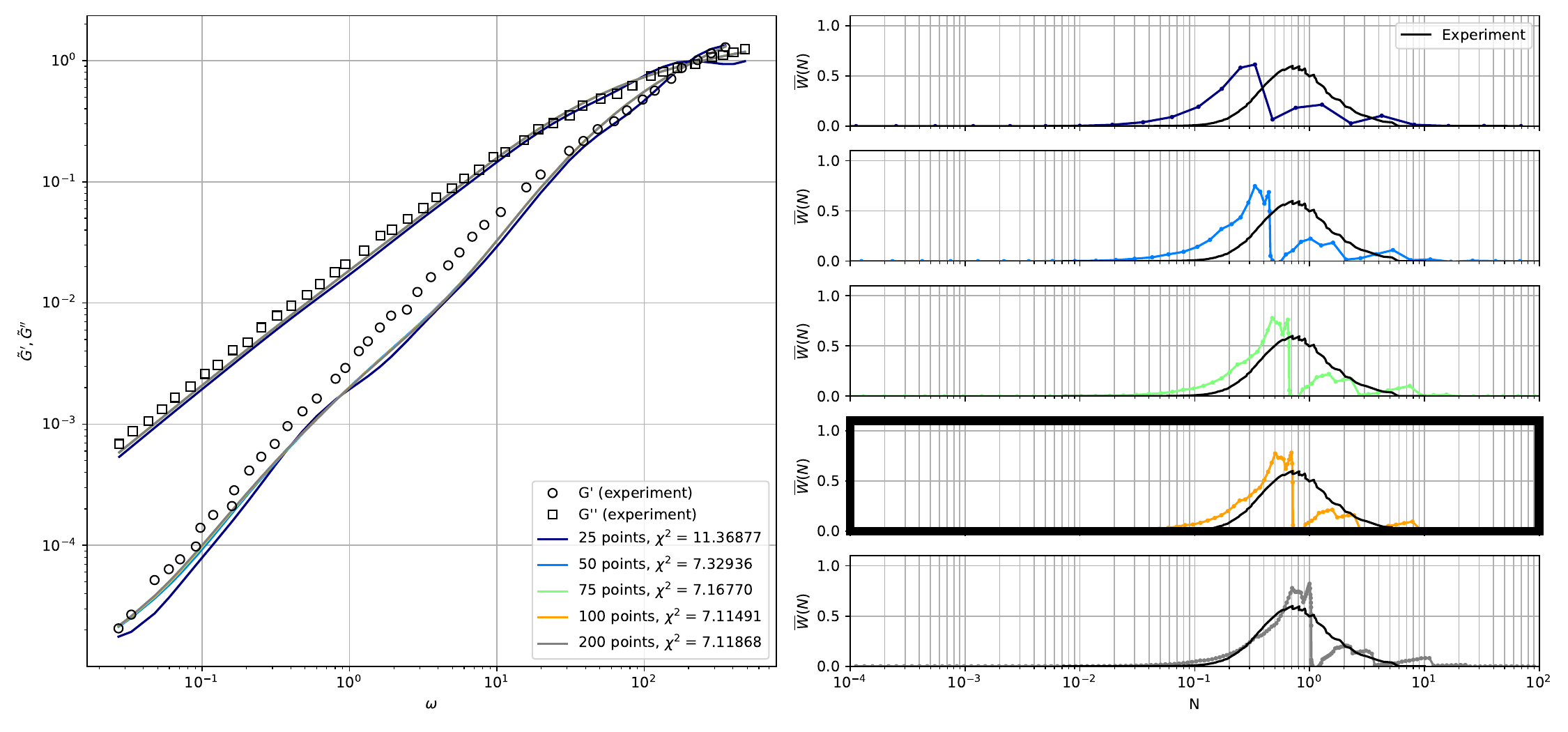}
    \caption{
        Inverse method results for PE3.
        Experimental moduli data is from Fig 9c and experimental MWD from Fig 1b3 of \cite{vanruymbeke2002evaluation}.
        The bold box indicates the optimal fit, based on the $\chi^2$ criterion, at $n = 100$ with parameters $\lambda = 1.0$, $\mu = 10.0$.
        Even though moduli data is well matched, predicted MWD is relatively nonsmooth.
        It is noted that increasing $\lambda$ causes a significant loss in the $\chi^2$ function, indicating that the solution is fundamentally nonsmooth.
    }
    \label{fig:PE3-final}
\end{figure}

The method is applied next to the PE3 dataset (\cref{fig:PE3-final}).
It is noted that PE3 has a narrower experimental MWD than PE1.
The most notable feature of the predicted MWD is the sharp jump in the middle, dividing the MWD into two distinct regions, indicating two distinct domains of viscoelastic response.
The rheological fit is still good, but not as close as is observed in other cases.
This may be attributed to the sharp discontinuity in the MWD, which causes a irreducible contribution from the curvature penalty.

\begin{figure}[!hbt]
    \centering
    \includegraphics[width=\linewidth]{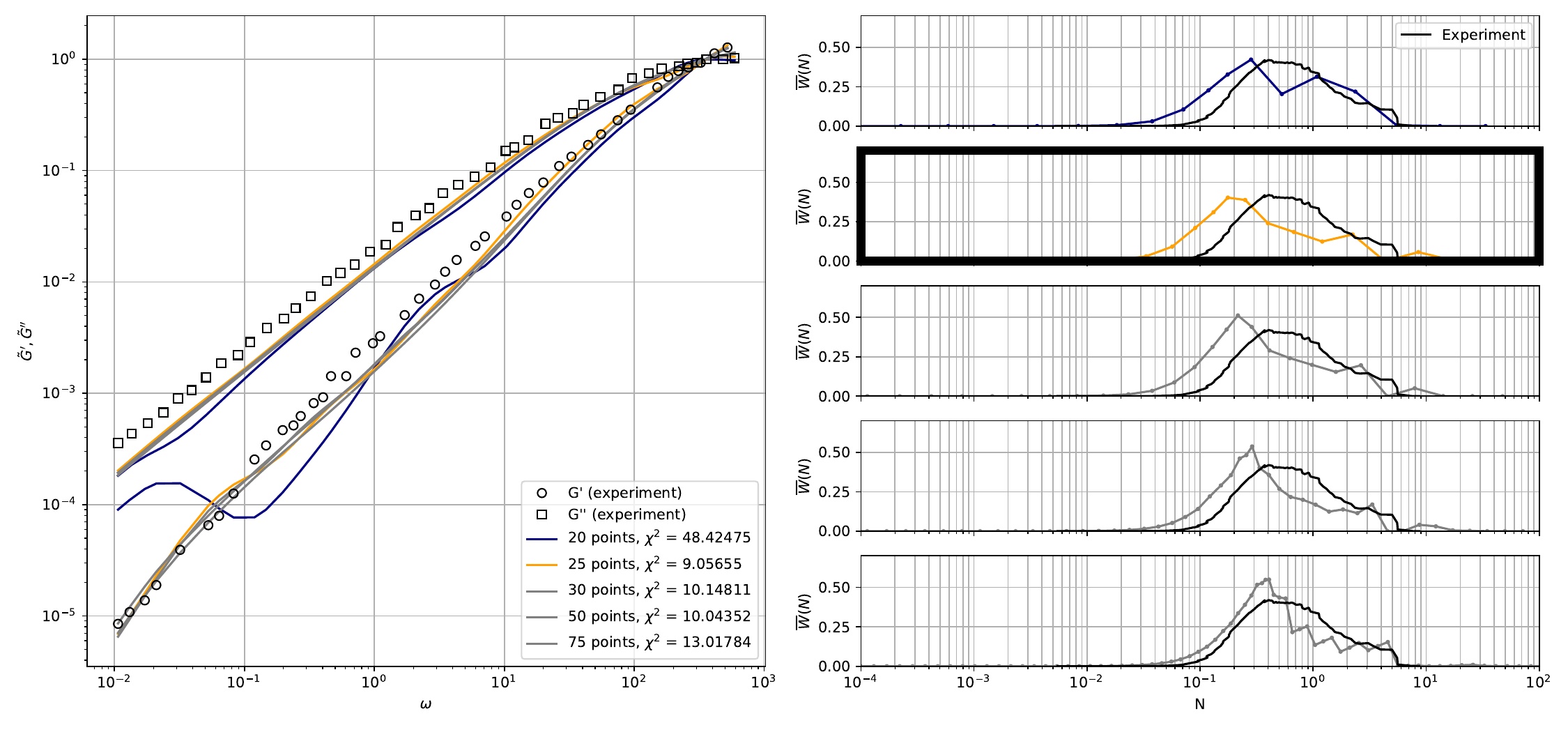}
    \caption{
    Inverse method results for PE4.
    Experimental moduli data is from Fig 9d and experimental MWD from Fig 1b4 of \cite{vanruymbeke2002evaluation}
    The bold box indicates the optimal fit based on the $\chi^2$ criterion, which is obtained at $n = 50$ with parameters $\lambda = 5.0$, $\mu = 5.0$.
    Notably, the shape of MWD remains the same with increasing $n$.
    }
    \label{fig:PE4-final}
\end{figure}

The inverse method is then applied to PE4 (\cref{fig:PE4-final}), which also has a narrower MWD range.
As with PE1 and PE3, a good rheological match is attained, especially at the crossover point.
The convergence criterion is then used to determine the optimal $n$ value, which in this case appears to be $n=25$ as the squared norm is minimized at this point.
This is highlighted to enforce that no prior knowledge of the MWD is used to determine the final prediction.
However, it is very interesting to note that, unlike other cases that exhibited overfitting, the predicted MWD steadily improves, with $n=75$ exhibiting a very close match to the experimental data.
With the addition of high-frequency rheological data, it is expected that this close match would have been identified at a lower value of $n$.

\subsection{Polystyrene}

Finally inverse method is applied to polystyrene (PS) data from literature.
Characterization of rheological and MWD data for both are taken from \textcite{guzman2005regularization} for polystyrene samples, which was commercially distributed by Atofina. 
It was originally reported that model parameters $\alpha = 3.60$, $\beta = 2.00$ work well with the double reptation model, without considering short-chain dynamics, and these parameters are used for the inverse method. 
Moreover, it was noted that the moduli behavior is indistinguishable up to $50 \times \omega_C$ when compared with including the Rouse dynamics. 
This implies that most of the information about the MWD is contained in the low-frequency domain, and that polystyrene is a good candidate for inverse modeling using the reptation approximation.

\begin{figure}
    \centering
    \includegraphics[width=\linewidth]{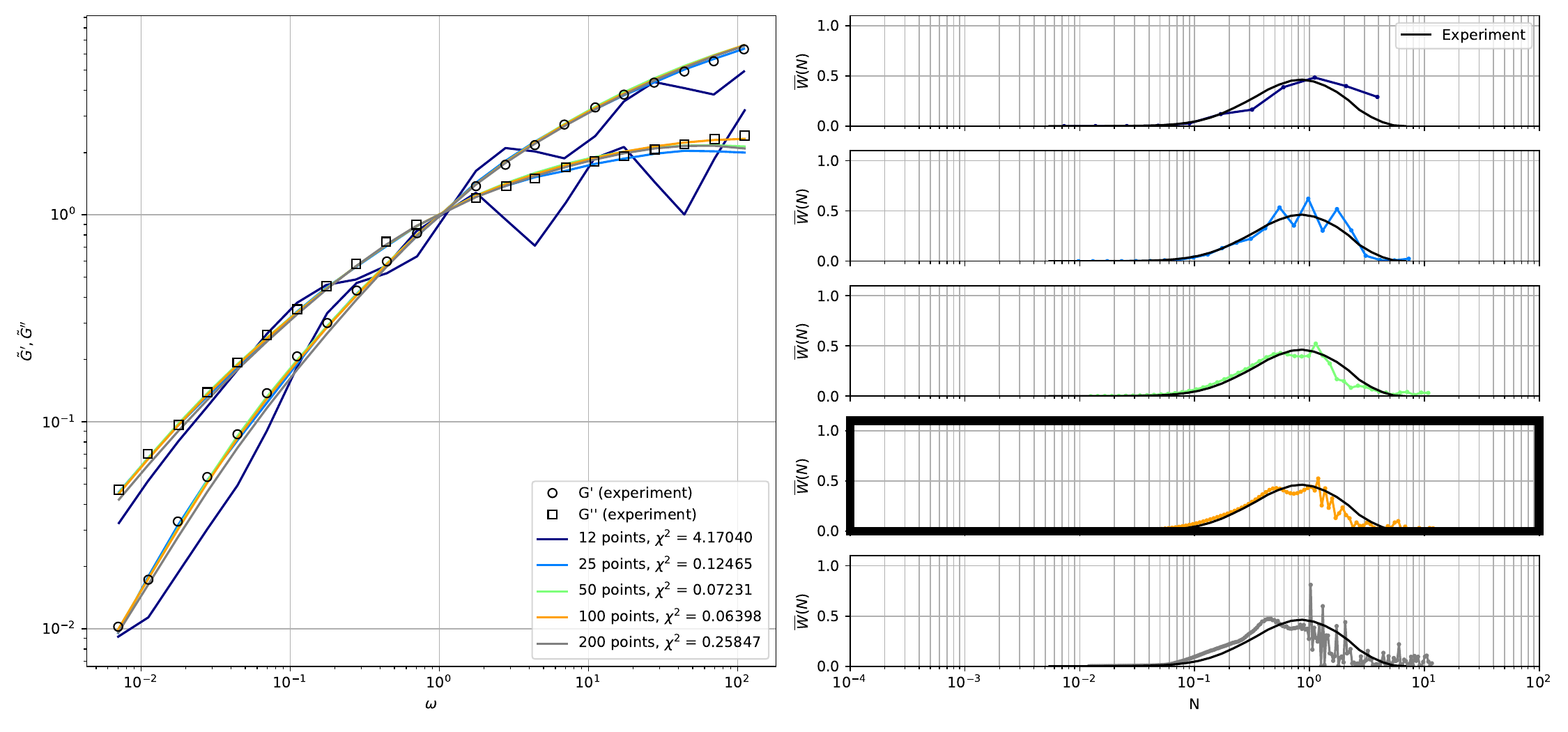}
    \caption{
    The inverse method results for PS.
    Experimental moduli data is from Fig 1 and experimental MWD from Fig. 5a of \cite{guzman2005regularization}.
    The bold box indicates the optimal fit to $\chi^2$ (local minima) which is obtained at $n = 100$ with parameters $\lambda = 3.0$, $\mu = 5.0$. 
    Increasing $n$ does creates problem for the optimizer and false peaks are observed. 
    }
    \label{fig:PS-final}
\end{figure}

The inverse method is applied to the literature data as with the previous samples (\cref{fig:PS-final}).
Using the minimum $\chi^2$ criterion, the n=100 case is selected as optimal, though it is apparent from visual inspection of the rheological data that all of the results (except for n=12) reasonably capture the moduli.
Increasing $n$ improves the accuracy in capturing both the moduli and MWD, though the unimodal distribution is slightly divided into two subtle yet persistent peaks. 
After $n \geq 100$, there is no significant improvement in moduli curve predictions, and overfitting begins to occur near the boundary of the $N$ range. 
Without knowledge of the experimental MWD, the $n=12$ case is sufficiently close to rheological data to yield a good result.
(It is noted that the PS results are very similar to PC-1 results in terms of the convergence of the solution, even though they are completely different polymer groups.)

\subsection{Discussion}

For all of the polymer samples considered, a reasonable match to the actual MWD was found.
The degree of the match depended on a variety of factors, including the range of rheological data available, the specific polymer species, the MWD of the polymer itself, and the degree to which each polymer followed the double reptation forward model.

It is important to emphasize that the method, in every case, produced a MWD that accurately reproduced the rheological data under the double reptation forward model.
Consequently, the output of the inverse method should be interpreted as providing the exact double-reptation-type polymer MWD that yields the rheological behavior.
Deviations from the experimental MWD, therefore, provide insight into how exactly each polymer differs from double reptation behavior.
For example the PC2 rheological behavior clearly exhibits a sharp kink in the G', G'' curves near the crossover point (\cref{fig:PC2-final}).
The MWD output of the method shows that a much sharper peak measured experimentally, with additional peaks to each side, yields an almost exact rheological model under double reptation.
From this, we may learn that this rheology may be driven by multiple modality, not necessarily in the MWD itself, but in the relaxation behavior.

\section{Conclusion}\label{sec:conclusion}

This work presents a generalized inverse method for predicting MWD from rheological data without restricting the functional form of the MWD. 
Unlike prior work, which generally restricts the MWD to a functional form characterized by a small number of degrees of freedom, this approach makes no assumptions about the shape of the MWD. 
This has the important advantage that no prior knowledge about the MWD is needed for the inverse method to work; in addition to capturing descriptors (such as average molecular weight and polydispersity), the method can now predict the overall shape of the MWD. 
By using penalty-method-based constraints, it is possible to convert the constrained optimization problem to an unconstrained problem, allowing for the use of powerful optimizers to solve efficiently.

The proposed inverse method is applied to a selection of polymers whose behavior is—to varying degrees—well-characterized by reptation-based rheological models. 
In cases where the polymer is very well-described by the forward model, the method is able to predict MWD with a high degree of accuracy. 
In cases where the polymer is not as well-described, or in cases where the rheological data is not provided over a sufficiently wide range, the method predicts MWDs inconsistent with experiment but consistent with rheological behavior under the forward model. 
While a better reptation model is needed for accurate predictions in such cases, the results still provide insight into deviations from reptation-based behavior.

An important observation is that the LM method routinely achieves a near-exact match to rheological data, yet predicted MWDs often differ from experimental results. 
These differences increase with higher MWD resolution, leading to overfitting without deterioration of rheological fits.
This demonstrates the non-uniqueness of optimal MWDs and the need for constraints. 
The LM method imposes smoothness constraints, stabilizing predictions but potentially limiting its ability to capture sharp peaks in MWDs. 
This highlights that (i) predicting exact MWD shapes from rheological data is theoretically impossible, and (ii) user-defined constraints aligned with expected results are essential.

We conclude by discussing limitations. 
First, as has been mentioned, this work focused on reptation-based models rather than the full spectrum of rheological theories.
The inverse method can in principle be applied to other models, but such an extension is beyond the scope of this study. 
Finally, it is noted that the present method introduces a smoothness requirement, and relies on a P(N)-based formulation that may require numerical differentiation when applied to arbitrary rheological models.

\section{Acknowledgments}

This work was supported by the US Department of Energy through the Los Alamos National Laboratory.
The Los Alamos National Laboratory is operated by Triad National Security, LLC, for the National Nuclear Security Administration of the U.S. Department of Energy (contract no. 89233218CNA000001).

\printbibliography

\end{document}